# Designing for Critical Algorithmic Literacies

Sayamindu Dasgupta[1,*]     Benjamin Mako Hill[2]

As pervasive data collection and powerful algorithms increasingly shape children's experience of the world and each other, their ability to interrogate computational algorithms has become crucially important. A growing body of work has attempted to articulate a set of "literacies" to describe the intellectual tools that children can use to understand, interrogate, and critique the algorithmic systems that shape their lives. Unfortunately, because many algorithms are invisible, only a small number of children develop the literacies required to critique these systems. How might designers support the development of critical algorithmic literacies? Based on our experience designing two data programming systems, we present four design principles that we argue can help children develop literacies that allow them to understand not only how algorithms they work, but also to critique and question them.

[1] University of North Carolina at Chapel Hill
[2] University of Washington

[*] Correspondence: Sayamindu Dasgupta <sayamindu@unc.edu>

## Introduction

As pervasive data collection and powerful algorithms increasingly shape children's experience, young people's ability to interrogate computational algorithms is increasingly important. A growing body of work has sought to identify the intellectual tools that children might use to understand, interrogate, and critique the powerful algorithmic systems and to advocate for equipping children with them. Although terminology is still diffuse, we call these intellectual tools that allow young people to understand and critique the algorithmic systems that affect their lives *critical algorithmic literacies*. Unfortunately, because many of these powerful algorithms are invisible, only a small number of young people develop these literacies. In this article, we describe how designers can build systems to support the development of critical algorithmic literacies in children.

Reflecting on extensive observation and design work in the Scratch online community over the last decade, we offer four design principles that we argue can support children in developing critical algorithmic literacies:

1. Enable connections to data

2. Create sandboxes for dangerous ideas
3. Adopt community-centered approaches
4. Support thick authenticity

Our first principle encourages designers to *enable connections to data* by offering children opportunities to engage directly in data analysis, especially with data sets that relate to the world that the children live, learn, and play in. The rationale for this principle is that in an increasingly data-driven world, understanding algorithms is deeply connected to understanding data. As children engage in data analysis in order to ask and answer their own questions or pursue their own interests, they create their own algorithms. Through this process, they can start to interrogate both their data and their algorithms.

Our second principle suggests that the development of critical algorithmic literacies can be supported by *creating sandboxes for dangerous ideas*. Algorithms are both powerful and risky. Our design work suggests that children can develop a deep understanding of both facts when they are allowed to create and experiment with algorithms using carefully designed toolkits. Because these toolkits entail giving learners the ability to "play with fire" in ways that might lead to negative outcomes, effective toolkit design needs to ensure that the possible dangers are managed and minimized. We use the metaphor of "sandboxes" to describe the goal of managing risk in this design process.

Our third principle suggests that designers should *adopt community-centered approaches* that allow designs to leverage community values that algorithms might challenge. Children belong to many overlapping communities and will typically share many of their communities' values. Algorithm are seen as problematic, by children and by society in general, when they violate these socially constituted values. A community-centered approach intentionally situates algorithms within communities with particular sets of shared values. Doing so makes the problematic nature of algorithms visible to learners who are likely to be aligned with community values that an algorithm violates or challenges.

Finally, we argue that *supporting thick authenticity*—a principle that applies to learning technology design in general—plays a crucial role in the development of critical algorithmic literacies. Authenticity in the context of fostering algorithmic or data literacies might mean engaging in activities that consider "real-world" data or scenarios.

Our paper is structured as follows. First, we describe the theoretical work that informs the way we conceptualize "critical data literacies" as well as the empirical and design work we have conducted that has informed our design principles. Next, we describe and situate the four design principles with detailed examples. Finally, we, discuss our principles' implications for future design work and conclude with a reflection on unanswered questions and future directions.

## Background

Our work draws from the literature on *constructionism*, a framework for learning and teaching that emphasizes contexts of learning "where the learner is consciously engaged in constructing a public entity, whether it's a sand castle on the beach or a theory of the universe" (Kafai 2006; Papert and Harel 1991:1). We are particularly inspired by Resnick and Silverman (2005) who provide a series of design principles for designing constructionist learning environments and toolkits based on reflections on their practice as designers. This article attempts to follow in Resnick and Silverman's footsteps by laying out design principles for critical algorithmic literacies.

We use the term "*algorithmic literacies*" to describe a subset of computational literacies as articulated by diSessa (2001) in his book, *Changing Minds: Computers, Learning, and Literacy*. diSessa suggests three broad pillars for literacy—material, mental or cognitive, and social. Material involves signs, representations, and so on. For language literacy, the material pillar might include alphabets, syntax, conventions of writing. For computational literacies, the material might involve user interface paradigms like spreadsheets or game genres, or modes of transmission like sharing on social media. The second pillar—mental or cognitive—represents the "coupling" (p. 8) of the material and what goes on inside learners' minds when interacting with the material. The final pillar—social—represents communities that form the basis of literacies. diSessa posits that the emergence of a given literacy is driven by "complex social forces of innovation, adoption, and interdependence" (p. 11).

More recently, Kafai, Proctor, and Lui (2019) has proposed a framework with three frames for understanding computational thinking: the cognitive, the situated, and the critical. They call for approaches to computational thinking that integrate *cognitive understanding* in the form of comprehension of computational concepts, *situated use* meaning that learning happens in contexts the learner cares about, and *critical engagement* to emphasize the importance of supporting the questioning of larger structures and processes behind the phenomenon being analyzed. These three frames can also be used in the context of computational literacies. In fact, one of the case studies used by Kafai et al. to illustrate their framework is framed around the concept of "critical data literacies" drawn from our work (Hautea, Dasgupta, and Hill 2017).

Our use of the term "critical" draws from Agre's (2014) idea of "critical technical practice" which ties critique and questioning to the practice of building and creation. In that sense, our goal is not merely knowledge about algorithms (e.g., what algorithms are) but an ability to engage in critique of algorithmic systems reflexively. Agre posits critical technical practice as requiring a "split identity—one foot planted in the craft work of design and the other foot planted in the reflexive work of critique" (p. 155). We recognize that as children engage with our toolkits, their design work combined with their reflection allows them to not only understand technical concepts around algorithms (what Agre describes as "esoteric terms"), but also evaluate their implications on society ("exoteric terms").

Finally, the notion of critical data literacies, as we have used before, is rooted in Freire's (Freire 1986) literacy methods. As we use it, the term was first proposed by Tygel and Kirsch (2016) who noted parallels between Freirean approaches to literacy education and the potential of models for developing data literacy. In suggesting approaches to big data literacy, D'Ignazio and Bhargava (2015) also build on Freire to posit that "[big data] literacy is not just about the acquisition of technical skills but the emancipation achieved through the literacy process" (p. 5). Relatedly, Lee and Soep (2016) have described their extensive body of work with youth-driven multimedia production at the "intersection of engineering and computational thinking on the one hand, and narrative production and critical pedagogy on the other" (p. 481) in terms of a *critical computational literacy*—a framework first developed by Lee and Garcia (2015) while studying youth from South Los Angeles creating animations and interactive games about socio-political issues in their community like racial profiling.

Our design principles are the result of design and empirical research around two systems we have developed and deployed over the last ten years: *Scratch Cloud Variables* and *Scratch Community Blocks*. Both tools were designed with constructionist framings of learning in mind toward supporting young people in learning about computational concepts related to data collection, processing, and analysis. Both tools also built upon and extended the Scratch programming language—a widely used block-based programming language for children (Resnick et al. 2009)—and were deployed in the Scratch online community where Scratch users share, comment on, and remix their Scratch projects (Monroy-Hernández and Resnick 2008).

The primary design goal of *Scratch Cloud Variables* was to give children the ability to collect, record, and analyze data within Scratch (Dasgupta 2013a). The primary goal of *Scratch Community Blocks* was to give children the ability to engage in analysis of their own social data directly (Dasgupta and Hill 2017). *Scratch Community Blocks* enabled this goal by allowing Scratch users the ability to access and analyze data from the Scratch online community website database. In deploying both systems, we found that granting children programmatic access to data led them to not only learn the techniques of data analysis but to question and critique data-driven algorithms.

## Design principles

Over the last decade, much of our research has focused on the design, deployment, and study of systems that seek to support constructionist learning around data and data-intensive algorithms. We distill lessons from this work into a four principles that we believe will be useful for designers interested in supporting young people to learn about data-driven computational techniques, as well as to question and to resist them.

### Principle 1: Enable connections to data

Our first principle suggests that algorithmic literacies can be supported by offering opportunities that enable children to write programs that interact with data that relate to the children's worlds. This principle stems from our experience with both *Scratch Cloud Variables* and *Scratch Community Blocks*. In both cases, we have found that even relatively simple connections to data from a programming toolkit enables scenarios where children ask questions about the algorithms that shape, store, and use information they create and care about.

We developed *Scratch Cloud Variables* as a part of the second generation of the Scratch programming language (Scratch 2.0). The system allowed Scratch users to store values in variables in ways that persists beyond the run-time of their program and is global in that everyone using their project would see the same data (Dasgupta 2013a). This support for persistent global data, combined with the fact that Scratch 2.0 projects were stored online and ran in a web-browser, allowed for functionality in Scratch projects such as global high-score lists, surveys, collaborative art projects, and more (figure 1).

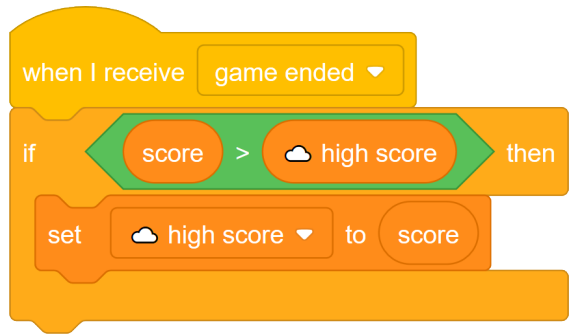

Figure 1: An example of a Scratch script using Scratch Cloud Variables. The script determines if the score of the concluded game is higher than the recorded high score. If so, the cloud variable `high score` (indicated by the cloud icon before the variable name) is updated.

During early testing of the system, children raised concerns about potential threats to privacy made possible by the system (Dasgupta 2013b). As a Scratch project could store data persistently, it was possible to create relatively simple Scratch code that would ask for someone's Scratch username (e.g., for a Scratch "guestbook" project) and store it indefinitely. The only way to remove the data, once stored, was to ask the creator of the project to do so. This example shows that even relatively simple connections with data open up possibilities that enable young people to think about questions of algorithmic surveillance and power.

We developed by *Scratch Community Blocks* by adding programming blocks representing programming primitives into Scratch that could access public metadata about projects and users in the Scratch online community database (Dasgupta and Hill 2017). An example of the system is shown in figure 2. For example, with *Scratch Community Blocks*, it was possible to create Scratch programs that would access how many times a Scratch project shared in the community had been viewed. Community-wide statistics such as total number of registered users in the community were also accessible programmatically through *Scratch Community Blocks*. These two sets of programming blocks were combined by a young Scratch user to make a project that would calculate what proportion of the broader Scratch community has viewed a given Scratch project.

Soon after this project was shared, through a series of comments on this particular project, Scratch community members realized that there was a difference between views and viewers. A single user may view a project multiple times in ways that cause the project's view count to increase. But only in some cases, as at the time the Scratch website counted views using an algorithm that both tried to count as many views as possible (e.g., from non-logged in users) so that creators of projects would see that their creations had an audience while also preventing community members from generating views synthetically (e.g., by repeatedly refreshing a project page).[1] Community members noted that one of the most popular projects on the site had a view count that exceeded the number of user accounts on the site. Through this process, users collectively realized that data is not objective, but requires *interpretation*, and that the process of data generation is shaped by decisions taken by others (Hautea et al. 2017).

---

[1] The first author of this article had implemented this particular algorithm for the Scratch website at that time.

Bowker (2005) has argued that "raw data is [...] an oxymoron" (p. 184). In a synonymous edited volume, Gitelman (2013) note that "data are imagined and enunciated against the seamlessness of phenomena" (p. 3). Often, this imagination and enunciation materializes in an algorithm that collects data, such as the viewership statistics of projects in Scratch. Enabling children to access that data through algorithms that they implemented using *Scratch Community Blocks* led them to discover illuminating patterns in data such as the fact that view-counts of popular projects were exceeding the total number of community members. This, in turn led children to attempt to reconstruct how data might have been imagined in the first place.

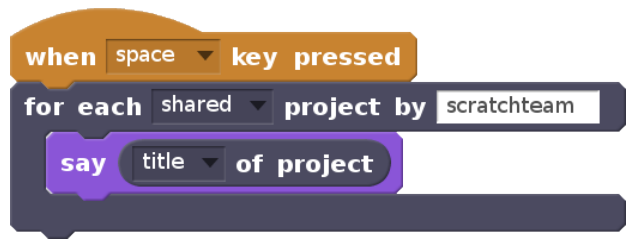

Figure 2: Example code using Scratch Community Blocks. The code iterates through all the projects shared by the user "scratchteam." In each iteration, the Scratch character being controlled by the code says the title of the project. Image from Dasgupta and Hill (2017).

Resnick and Silverman (2005) posit that "a little bit of programming goes a long way" in that children can combine relatively simple and limited programming constructs toward a broad range of creative outcomes. In our work, we see a similar phenomenon emerge where simple programming constructs, combined with data in straightforward ways, enable children to uncover structures and assumptions in algorithmic systems. This process allows children to raise questions and engage in conversations about algorithmic data collection.

**Principle 2: Create sandboxes for dangerous ideas**

Our second principle suggests that the development of critical algorithmic literacies can be supported through the creation of sandboxes for dangerous ideas. Algorithms are dangerous in that although they are implemented in computational systems that operate on abstract and incomplete models of the world, they have powerful, unanticipated, and often negative consequences (Smith 1985). For example, the algorithm behind a real-estate search tool may allow the user to filter houses that are for sale by school rating but are unlikely to take the history of underfunding of schools in African American and low income neighborhoods into account. In this way, the search algorithm might unintentionally become a way for potential house-buyers to filter for affluent predominantly white neighborhoods (Noble 2018:167).

With the deployment of *Scratch Community Blocks*, metadata about user accounts such as number of followers and number of projects were made programmatically accessible. These numbers can be used as proxies for measures of experience in that more projects or more followers is an indication of more experience with Scratch—but both are far from perfect measures. Although restricting interaction with a project to more experienced users might be an attractive feature to some Scratch users, using these measures as a gate-keeping mechanism can be discriminatory for newcomers. This was exactly the concern raised by an young member of the Scratch community who noted that the algorithm to carry out such discrimination is trivial using *Scratch Community Blocks*—a single `if` statement. The young user noted that algorithmic systems can be dangerous in that they can they can enable surveillance, discrimination, and more.

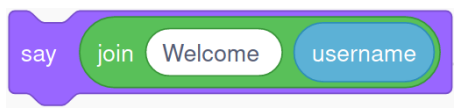

Figure 3: The `username` block introduced in Scratch 2.0. The block "reports" or returns the username of the person viewing the project, or blank if the viewer is not logged in.

In this sense, algorithms reflect dangerous ideas. The notion of engaging with and exploring dangerous ideas is not new in education: problematic theories are studied as a part of understanding history; discriminatory scenarios are analyzed as a part of engaging with the idea of justice; and potentially physically dangerous experiments are carried out in school and college chemistry laboratories. Although these activities all represent different types of danger, the pedagogical activities around them typically incorporate appropriate safety mechanisms. For the pedagogy of fields like chemistry, this is a topic of ongoing research and study (Alaimo et al. 2010).

An example from our own work that led us to consider the notion of dangerous ideas is a feature introduced in the second generation of Scratch—an `username` block that "reports" the user name of the viewer of the project if they are logged in (figure 3). The `username` block allowed for new types of surveillance in that it made it much easier to know who had accessed a given project. As designers, we were also concerned that that the block can be used for discriminatory purposes with a Scratch project (e.g., by disallowing certain usernames from playing a Scratch game), or to evade moderators (e.g., to have a Scratch project behave in a specific way only for known moderators in the community). On the other hand, it also made new conversations around surveillance, discrimination, data, and algorithms possible. Achieving a balance between enabling exploration of dangerous ideas and safety is not easy. This is especially the case among historically marginalized groups in STEM learning who may be more vulnerable to discrimination and surveillance.

We only began to seriously adding consider the feature because usernames in Scratch are, by community policy, not directly tied to identities in the real world. As a result, the consequences of surveillance in Scratch would be less serious than surveillance of email accounts or other social media accounts. We also considered a range of approaches to make the username block safer. Many of these were technical. For example, an initial prototype of the block reported back an alpha-numeric value that would remain consistent for a given user accessing a given project over time so that a user interacting with the project could not be identified by username.[2] Because this idea was difficult to explain, we did not adopt this approach. As an added measure, the Scratch project viewing interface was modified so that it warned users about the existence of the block in a given project before they ran it and encouraged users to log out of Scratch if they wished to avoid potential tracking.

---

[2]Internally, this alphanumeric value was being generated by one-way cryptographic hash of the username, project id, and a salt that was global to the website.

We also deployed the `username` block with considerable caution, carefully monitored its usage, and were ready to roll back the feature. The Scratch community has a complex and extensive moderation and governance infrastructure that has been described by Lombana Bermúdez (2017) as a combination of "proactive and reactive moderation [...] with the cultivation of socially beneficial norms and a sense of community" (p. 35). We felt that these structures had the potential to prevent and mitigate misuse of the feature. In the design phase of the `username` block, we engaged in a number of conversations with the Scratch's moderation team and with children. These conversations continued after the feature's launch, so that we could gain an understanding of how the new block was being used by the broader community and adapt the system if needed.

A metaphor to describe our approach of trying to strike a balance between "dangerous ideas" and the risks associated with such ideas is that of "sandboxes." Just as a sandbox allows children to play with earth in ways that may not be suitable for the rest of the playground, our design consists of creating well-defined boundaries. The metaphor of a sandbox is common in the field of computer security where untrusted applications are said to run in a "sandbox" isolated from unneeded resources and other programs (Schreuders, McGill, and Payne 2013). For example, a mobile phone sandboxing systems ensures that an app that does not need access to the camera does not have access to it. In the case of the `username` block, we spent several design iterations working with children and community moderators to ensure that there were enough safety "walls" (e.g., warning message in projects that use the block) before we felt that we had achieved a balance between encouraging explorations and safety. In computer and information security pedagogy, the use of sandboxes to allow learners to experiment with software vulnerabilities is an established practice (Du and Wang 2008). Computer security researchers and educators have asked students to construct speculative fiction to engage with dangerous ideas and to imagine these ideas' impact on society (Kohno and Johnson 2011). Our experience suggests that a similar approach may work for critical algorithmic literacies as well.

**Principle 3: Adopt community-centered approaches**

Our third principle suggests that designers should incorporate community-centered approaches that allow a design to leverage existing community values that an algorithm might change or challenge. This principle reflects increasing recognition of the importance of centering values in computing learning. For example, in a keynote presentation to the 2012 ACM SIGCSE Technical Symposium, Abelson (2012) called for a focus on "computational values," which Abelson defined as commitments that "empower people in the digital world," and which he argued are "central to the flowering of computing as an intellectual endeavor." Justice, respect for privacy, and non-discrimination are examples of such values.

Prior work in Human-Computer Interaction literature has argued that values are "something to be discovered" in the context of a given community (Le Dantec, Poole, and Wyche 2009:1145). In turn, values can also influence the sense of identity of a learner within their communities. In recent work in the Learning Sciences, Vakil (2020) has proposed the phrase "disciplinary values interpretation" to describe how learners seek to understand what a discipline being studied "is 'all about,' and what it might mean for them to be a part of it as they begin to imagine their future academic, career, and life goals" (p. 7). Vakil has also called for more understanding of, and attention to, "adolescents' political selves and identities, and how these identities become intertwined with learning processes" (p. 22).

In our work, we have seen the dynamic described by Vakil play out as young people evaluate technological possibilities in terms of their values. For example, we saw that children using *Scratch Community Blocks* questioned algorithms by describing algorithmic outcomes as in conflict with the collective value of the Scratch community. Multiple community members expressed concerned about *Scratch Community Blocks* enabling projects that rank community members based on the number of followers, and pointed out that this might shift the values of the Scratch community from celebrating creativity and expression to an emphasis on popularity. Similarly, the child using *Scratch Community Blocks* who pointed out that code using the new system could be used to block newcomers from projects thought this was problematic because inclusivity is a core value of the Scratch community and the algorithmic discrimination that they correctly identified was made possible by the system stood in contrast to this value.

One challenge with systems that enable possibilities that go against established community values is the fact that unsocialized newcomers will frequently not share their new community's values. Zittrain (2006) noted this as a challenge with "generative" systems and platforms where the outcomes made possible by the system include both positive and negative outcomes. With *Scratch Cloud Variables*, we recognized this issue and implemented a system where the *Scratch Cloud Variables* feature would only be made available to users who have been active in the community for some time (Dasgupta and Hill 2018). By only granting access to the dangerous feature to users likely to have been socialized, we reasoned that newcomers would be allowed to learn Scratch's community values before being given access to features that enabled them to flout them.[3]

Our experience suggests that critical approaches to algorithms are driven by the values of the communities in which algorithms are enacted. Of course, communities vary in scope and character and can range from groups of friends, to families, classrooms, and entire nations. In a 2014 report produced by the Executive Office of the President of the United States of America, values enshrined in the legal structure of the United States were invoked when it was stated that "big data technologies" have the "potential to eclipse longstanding civil rights protections in how personal information is used in housing, credit, employment, health, education, and the marketplace" (Podesta 2014:3).

Of course, not all values are aligned with outcomes that we as educators want. Values frequently conflict with each other and widely shared values can sometimes be problematic. Moreover, ways of imagining a specific value can overwhelm alternatives in ways that are described by by Benjamin (2019) as a "master narrative" (p. 134). For example, Philip, Schuler-Brown, and Way (2013) draw from their classroom experience in a U.S. public school setting to describe how the underlying assumption among students debating "big data" and its implications was "that students, particularly urban youth of color, would academically, socially, occupationally, or politically benefit simply by virtue of exposure to big data and new technologies" (p.117). Philip et al. explain that the design of their new curriculum did not take into account the relative lack of opportunities for students of color leading to an overwhelm-ing focus on one particular framing of big data technology as an equalizer. This example serves as a warning for designers to carefully interrogate a range of values before designing.

---

[3] A second part of the reasoning is that if someone used *Scratch Cloud Variables* in a problematic way, they might get banned and lose access to the account into which they had poured substantial time and resources.

**Principle 4: Support thick authenticity**

Finally, we argue that thick authenticity—a principle that applies to learning technology design in general—plays a crucial role in the development of critical algorithmic literacies. Authenticity is a complicated concept in the context of learning. Although it is common to encounter terms related to the "real world" and "real world problems" in popular and scholarly discourse on education, degrees and dimensions of "realness" vary enormously. For example, a learning exercise might involve a fictional scenario where a problem is real but a situation is not (Petraglia 1998).

"Realness", or authenticity is determined ultimately by the learners and "real" learning experiences, problems, and metaphors may be unfamiliar to learners because of cultural differences or differences in interest. An example based on draws from a pack of playing cards may hinder the learning experience for students of probability who have never played cards. As corollaries, stronger forms of authenticity emerge when learners have more say in the design and direction of their learning activities and higher degrees of authenticity are associated with better learning outcomes. For example, while teaching Maori schoolchildren English, Ashton-Warner (1986) found that a compelling strategy to engage her students was to ask them to write about themselves, about their own stories, in their own words—a process she called "organic writing".

Question of authenticity are likely to be relevant to the type of engagement necessary to support the development of algorithmic literacies in children. For example, a baseball analytics algorithm may generate critical engagement when the learner is a young baseball fan who is going to poke holes in the assumptions made by the algorithm. The same algorithm would likely elicit a lukewarm response from someone without an interest in baseball, at best. To most learners outside of the United States, learning activities that involve baseball are not suitable at all. In our design work, we have drawn inspiration from Ashton-Warner and asked what "organic writing" might look like for developing critical algorithmic literacies? (Dasgupta 2016)

We have also drawn from Shaffer and Resnick (1999) who describe "thick authenticity" as:

> [...] activities that are personally meaningful, connected to important and interesting aspects of the world beyond the classroom, grounded in a systematic approach to thinking about problems and issues, and which provide for evaluation that is meaningfully related to the topics and methods being studied.

In our work with *Scratch Community Blocks*, children using the system engaged with complex ideas about power and algorithms because the data that they were accessing, and the algorithms that they were designing, reflected their experiences, friends, and community in Scratch. If the same interface within Scratch had provided access to nearly any other data source, it would have been less effective as promoting algorithmic literacy among the community of Scratch users to whom our system was deployed. In that most children do not use Scratch, the effectiveness of our systems is likely to be limited among most children.

That said, other contexts might present similarly promising opportunities. For example, families' interactions are increasingly shaped by algorithms and data inherent in "smart home" technologies. Although it is still less common among children, a range of individuals are increasingly collecting data about aspects of their personal lives through quantified-self approaches (Lee 2013). In that algorithms are increasingly prevalent in a range of contexts, there is an increasingly wide range settings offering rich opportunities for the promotion of algorithmic literacies through thick authenticity.

## Discussion

In our own design experiences spanning many iterations, we encountered numerous tensions and open questions in terms of how to best engage the broadest possible set of Scratch community members in critiquing algorithms while making sure that the Scratch community values and ethos remain intact. The evidence emerging from our work suggests that there may be certain design principles—presented in this article—for computational construction kits that support the development of a range of critical algorithmic literacies. Our four design principles reflect a broader perspective that young learners should go beyond simply observing algorithmic systems and be given opportunities to *create* algorithmic systems of their own. We argue that when youth take advantage of these opportunities, they will often question algorithms in meaningful ways. In empirical work we have conducted, we employed grounded theory (Charmaz 2006) to analyze the discussions, comments, and activities of children engaging in creative design activities using *Scratch Community Block* (Hautea et al. 2017). Most of the examples we identified of children questioning algorithmic systems emerged from the process of active creation with toolkits.

That said, there may be other ways to effectively engage young people in understanding and questioning algorithmic systems. For example, approaches such as co-designed games have been found to yield promising results for engaging children in understanding notions of privacy (Kumar et al. 2018). Similar approaches may emerge toward other aspects of critiquing data and algorithm driven systems as well.

Our work is limited in that it has so far focused on individual learners and case studies. We have yet to conduct any systematic study of outcomes around critical algorithmic literacies in learning environments where our principles have been put into practice. We offer our principles in the hopes that other designers of educational technologies and experiences will build on our work and contribute to the larger project of fostering critical algorithmic literacies in young people.

## Conclusion

In their paper on designing construction kits for children, Resnick and Silverman (2005) present their final design principle as "iterate, iterate—then iterate again." They conclude by stating that this applies to their principles as well. Our four principles are no exceptions to this excellent advice. Going forward, we intend to keep iterating on our principles, taking them apart, putting them back together, and changing them. We offer our principles with humility and a sincere desire to work toward the dual goals of supporting young people in understanding the algorithmic systems that are increasingly shaping their worlds as well as to do what we can to give them the intellectual tools to questioning and resist them.

## Acknowledgments

Many of the projects referred to in this article were financially supported by the US National Science Foundation. We would also like to acknowledge feedback and support from Mitchel Resnick, Natalie Rusk, Hal Abelson, Brian Silverman, Amos Blanton, and other members of the Scratch team. Finally, none of this work would have been possible without the children who generously tried out our new technologies, gave us feedback, and inspired us in multiple ways with their ingenuity and kindness.

# References

Abelson, Hal. 2012. "From Computational Thinking to Computational Values." Pp. 239–40 in *Proceedings of the 43rd ACM Technical Symposium on Computer Science Education, SIGCSE '12*. New York, NY, USA: ACM.

Agre, Philip E. 2014. "Toward a Critical Technical Practice: Lessons Learned in Trying to Reform AI." Pp. 131–57 in *Social Science, Technical Systems, and Cooperative Work: Beyond the Great Divide.*, edited by G. Bowker, S. L. Star, L. Gasser, and W. Turner. Hoboken, NJ: Taylor; Francis.

Alaimo, Peter J., Joseph M. Langenhan, Martha J. Tanner, and Scott M. Ferrenberg. 2010. "Safety Teams: An Approach to Engage Students in Laboratory Safety." *Journal of Chemical Education* 87(8):856–61.

Ashton-Warner, Sylvia. 1986. *Teacher*. Simon; Schuster.

Benjamin, Ruha. 2019. *Race After Technology: Abolitionist Tools for the New Jim Code*. Medford, MA: Polity.

Bowker, Geoffrey C. 2005. *Memory Practices in the Sciences*. Cambridge, Mass: MIT Press.

Charmaz, Kathy. 2006. *Constructing Grounded Theory: A Practical Guide Through Qualitative Analysis*. London, UK: Sage Publications.

Dasgupta, Sayamindu. 2013a. "From Surveys to Collaborative Art: Enabling Children to Program with Online Data." Pp. 28–35 in *Proceedings of the 12th International Conference on Interaction Design and Children (IDC '13)*. New York, NY: ACM.

Dasgupta, Sayamindu. 2013b. "Surveys, Collaborative Art and Virtual Currencies: Children Programming with Online Data." *International Journal of Child-Computer Interaction* 1(3–4):88–98.

Dasgupta, Sayamindu. 2016. "Children as Data Scientists : Explorations in Creating, Thinking, and Learning with Data." Thesis, Massachusetts Institute of Technology.

Dasgupta, Sayamindu and Benjamin Mako Hill. 2017. "Scratch Community Blocks: Supporting Children as Data Scientists." Pp. 3620–31 in *Proceedings of the 2017 CHI Conference on Human Factors in Computing Systems (CHI '17)*. New York, New York: ACM Press.

Dasgupta, Sayamindu and Benjamin Mako Hill. 2018. "How 'Wide Walls' Can Increase Engagement: Evidence from a Natural Experiment in Scratch." Pp. 361:1–361:11 in *Proceedings of the 2018 CHI Conference on Human Factors in Computing Systems (CHI '18)*. New York, New York: ACM.

D'Ignazio, Catherine and Rahul Bhargava. 2015. "Approaches to Building Big Data Literacy." in *Proceedings of the Bloomberg Data for Good Exchange Conference 2015*. New York, N.Y.

diSessa, Andrea A. 2001. *Changing Minds: Computers, Learning, and Literacy*. Cambridge, MA: MIT Press.

Du, Wenliang and Ronghua Wang. 2008. "SEED: A Suite of Instructional Laboratories for Computer Security Education." *Journal on Educational Resources in Computing* 8(1):1–24.

Freire, Paulo. 1986. *Pedagogy of the Oppressed*. New York: Continuum.

Gitelman, Lisa, ed. 2013. *"Raw Data" Is an Oxymoron*. Cambridge, Massachusetts ; London, England: The MIT Press.

Hautea, Samantha, Sayamindu Dasgupta, and Benjamin Mako Hill. 2017. "Youth Perspectives on Critical Data Literacies." Pp. 919–30 in *Proceedings of the 2017 CHI Conference on Human Factors in Computing Systems, CHI '17*. New York, NY, USA: ACM.

Kafai, Yasmin B. 2006. "Constructionism." Pp. 35–46 in *The Cambridge Handbook of the Learning Sciences*, edited by K. R. Sawyer. Cambridge, UK: Cambridge University Press.

Kafai, Yasmin, Chris Proctor, and Debora Lui. 2019. "From Theory Bias to Theory Dialogue: Embracing Cognitive, Situated, and Critical Framings of Computational Thinking in K-12 CS Education." Pp. 101–9 in *Proceedings of the 2019 ACM Conference on International Computing Education Research, ICER '19*. New York, NY, USA: ACM.

Kohno, Tadayoshi and Brian D. Johnson. 2011. "Science Fiction Prototyping and Security Education: Cultivating Contextual and Societal Thinking in Computer Security Education and Beyond." P. 9 in *Proceedings of the 42nd ACM technical symposium on Computer science education - SIGCSE '11*. Dallas, TX, USA: ACM Press.

Kumar, Priya, Jessica Vitak, Marshini Chetty, Tamara L. Clegg, Jonathan Yang, Brenna McNally, and Elizabeth Bonsignore. 2018. "Co-Designing Online Privacy-Related Games and Stories with Children." Pp. 67–79 in *Proceedings of the 17th ACM Conference on Interaction Design and Children - IDC '18*. Trondheim, Norway: ACM Press.

Le Dantec, Christopher A., Erika Shehan Poole, and Susan P. Wyche. 2009. "Values as Lived Experience: Evolving Value Sensitive Design in Support of Value Discovery." P. 1141 in *Proceedings of the 27th international conference on Human factors in computing systems - CHI 09*. Boston, MA, USA: ACM Press.

Lee, Clifford H. and Antero D. Garcia. 2015. "'I Want Them to Feel the Fear...': Critical Computational Literacy as the New Multimodal Composition." Pp. 2196–2211 in *Gamification: Concepts, Methodologies, Tools, and Applications*. Hershey, PA, USA: IGI Global.

Lee, Clifford H. and Elisabeth Soep. 2016. "None but Ourselves Can Free Our Minds: Critical Computational Literacy as a Pedagogy of Resistance." *Equity & Excellence in Education* 49(4):480–92.

Lee, Victor R. 2013. "The Quantified Self (QS) Movement and Some Emerging Opportunities for the Educational Technology Field." *Educational Technology* 53(6):39–42.

Lombana Bermúdez, Andrés. 2017. "Moderation and Sense of Community in a Youth-Oriented Online Platform: Scratch's Governance Strategy for Addressing Harmful Speech." in *Perspectives on Harmful Speech Online*. Cambridge: Berkman Klein Center for Internet & Society Research Publication.

Monroy-Hernández, Andrés and Mitchel Resnick. 2008. "Empowering Kids to Create and Share Programmable Media." *Interactions* 15(2):50–53.

Noble, Safiya Umoja. 2018. *Algorithms of Oppression: How Search Engines Reinforce Racism*. New York, NY: New York University Press.

Papert, Seymour and Idit Harel. 1991. "Situating Constructionism." Pp. 1–11 in *Constructionism*. Vol. 36. New York, NY, US: Ablex Publishing.

Petraglia, Joseph. 1998. *Reality by Design: The Rhetoric and Technology of Authenticity in Education*. Hoboken: Lawrence Erlbaum Associates.

Philip, Thomas M., Sarah Schuler-Brown, and Winmar Way. 2013. "A Framework for Learning About Big Data with Mobile Technologies for Democratic Participation: Possibilities, Limitations, and Unanticipated Obstacles." *Technology, Knowledge and Learning* 18(3):103–20.

Podesta, John. 2014. *Big Data: Seizing Opportunities, Preserving Values*. Executive Office of the President, United States of America.

Resnick, Mitchel, John Maloney, Andrés Monroy-Hernández, Natalie Rusk, Evelyn Eastmond, Karen Brennan, Amon Millner, Eric Rosenbaum, Jay Silver, Brian Silverman, and Yasmin Kafai. 2009. "Scratch: Programming for All." *Communications of the ACM* 52(11):60–67.

Resnick, Mitchel and Brian Silverman. 2005. "Some Reflections on Designing Construction Kits for Kids." Pp. 117–22 in *Proceedings of the 2005 Conference on Interaction Design and Children*, *IDC '05*. New York, NY, USA: ACM.

Schreuders, Z. Cliffe, Tanya McGill, and Christian Payne. 2013. "The State of the Art of Application Restrictions and Sandboxes: A Survey of Application-Oriented Access Controls and Their Shortfalls." *Computers & Security* 32:219–41.

Shaffer, David W. and Mitchel Resnick. 1999. ""Thick" Authenticity: New Media and Authentic Learning." *Journal of Interactive Learning Research* 10(2):195–215.

Smith, Brian Cantwell. 1985. "The Limits of Correctness." *ACM SIGCAS Computers and Society* 14,15(1,2,3,4):18–26.

Tygel, Alan and Rosana Kirsch. 2016. "Contributions of Paulo Freire for a Critical Data Literacy: A Popular Education Approach." *The Journal of Community Informatics* 12(3).

Vakil, Sepehr. 2020. " 'I've Always Been Scared That Someday I'm Going to Sell Out': Exploring the Relationship Between Political Identity and Learning in Computer Science Education." *Cognition and Instruction* 1–29.

Zittrain, Jonathan L. 2006. "The Generative Internet." *Harvard Law Review* 119(7):1974–2040.